\documentclass[twocolumn,10pt]{article}

\usepackage[a4paper, left=18mm, right=18mm, top=25mm, bottom=25mm, columnsep=20pt]{geometry}

\usepackage{amsmath}    
\usepackage{amssymb}    

\usepackage[T1]{fontenc}
\usepackage[utf8]{inputenc}   
\usepackage{newtxtext}
\usepackage{newtxmath}
\usepackage{microtype}

\usepackage{multirow}
\usepackage{booktabs}
\usepackage{graphicx}
\usepackage{xcolor}
\usepackage[authoryear, round]{natbib}
\usepackage[font=small, labelfont=bf, labelsep=period]{caption}

\usepackage{fancyhdr}
\newcommand{\shorttitle}{Probabilistic Deep Learning for Drought Forecasting}
\newcommand{\shortauthors}{Funk et al. (2026)}
\fancypagestyle{plain}{\fancyhf{}\fancyfoot[C]{\small\thepage}%
  }

\definecolor{preprintblue}{RGB}{20,60,130}
\usepackage[colorlinks=true, allcolors=preprintblue]{hyperref}
\usepackage{cleveref}

\newcommand{\orcid}[1]{\href{https://orcid.org/#1}{\includegraphics[height=8pt]{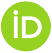}}}
\newcommand{\cross}[1][0.5pt]{\ooalign{%
  \rule[0.7ex]{0.7ex}{#1}\cr
  \hss\rule{#1}{.5em}\hss\cr}}

\begin{document}

\twocolumn[{%
\begin{center}
    {\LARGE\bfseries Probabilistic Deep Learning for Drought Forecasting:\\[2pt]
     The Role of Internal Climate Variability\par}

    \vspace{7mm}

    {\normalsize
    Henri Funk$^{1, 2,\cross, *}$\,\orcid{0009-0007-0949-8385},
    Cornelia Gruber$^{1,\cross}$\,\orcid{0009-0002-0657-3558},
    Göran Kauermann$^1$\,\orcid{0000-0003-0742-7835},\\[2pt]
    Helmut Küchenhoff$^1$\,\orcid{0000-0002-6372-2487},
    and Magdalena Mittermeier$^2$\,\orcid{0000-0002-8668-281X}
    \par}

    \vspace{3mm}

    {\small
    $^1$Department of Statistics, LMU Munich, Munich, Germany\\
    $^2$Department of Geography, LMU Munich, Munich, Germany\\[1.5mm]
    $^{\cross}$Authors who share equal contribution.\quad
    $^*$Author to whom any correspondence should be addressed. E-mail: \href{mailto:H.Funk@lmu.de}{H.Funk@lmu.de}
    \par}

    \vspace{6mm}

    \begin{minipage}{0.9\textwidth}
    {\color{black!40}\hrule height 0.4pt}
    \vspace{2.5mm}
    \small
    \noindent\textbf{Abstract.}
    Predicting drought risk is essential for anticipating impacts on water resources, agriculture, ecosystems, and climate adaptation planning. Yet drought forecasts remain uncertain because variability can substantially alter regional precipitation and evaporative demand. Treating this variability as unstructured noise ignores the fact that internal variability has spatial, seasonal, and temporal structure and thus contains information that can be used to improve drought forecasting. We propose a deep-learning-based forecasting framework for European drought prediction and extend it with an uncertainty-aware drought bound that explicitly incorporates internal forecast variability from a large climate model ensemble. This bound represents a physically plausible lower-tail trajectory of future drought conditions and marks how severe drought could plausibly become under an unfavourable realisation of internal variability, giving adaptation planning a conservative, risk-averse reference. We compare the proposed bound with a lower bound derived from reanalysis data only and show that our proposed ensemble-informed bound is better calibrated across most regions and seasons. This is specifically true during anomalously dry conditions, when historical reanalysis alone underestimates lower-tail drought risk. Our results show that internal variability should be treated as a forecast quantity in its own right. More broadly, large ensembles provide a practical way to transfer physically plausible climate variability into machine-learning drought forecasts, yielding risk-aware bounds that are more informative for drought assessment under shifting climate conditions.

    \vspace{2.5mm}
    \vspace{2.5mm}
    {\color{black!40}\hrule height 0.4pt}
    \end{minipage}
\end{center}
\vspace{8mm}
}]
\thispagestyle{plain}

\section{Introduction}

In 2022 and 2023, Europe faced severe drought conditions, driven by complex local thermodynamic and large-scale circulation anomalies that are expected to become more frequent in the future \citep{herrera2023, Toreti2023, bevacqua2024}.
Droughts are a recurrent hydrometeorological hazard, with direct implications for agriculture, water management, and related planning decisions.
Reliable drought information can therefore support risk assessment and early warning efforts in these sectors \citep{schuldt2020, bastos2020, bakke2020}.
At the same time, drought forecasting remains intrinsically difficult due to coupled land-atmosphere processes and large-scale circulation patterns whose regional variability varies substantially across seasons and regions \citep{Hao2018, prodhan2022}.
These complex dependencies place subseasonal forecasting in what has been called the \textit{predictability desert} \citep{richter2025, guan2026}, a scope in which uncertainty is inherently difficult to quantify and variability is routinely underestimated.

A central reason for this difficulty is natural variability.
Even in the absence of changes in external forcing, interactions within the climate system generate fluctuations in precipitation and temperature that limit predictability at regional scales \citep{hawkins2009, hawkins2011, deser2010}.
This variability can obscure or amplify forced signals and thereby sets a practical limit on the skill of drought forecasting \citep{lorenz1963, palmer2000}.
Therefore, forecasting alone is insufficient for drought risk assessment.
Forecasting systems need uncertainty information that reflects the full range of plausible hydrometeorological outcomes.

\begin{figure*}[t]
    \centering
    \includegraphics[width=\linewidth]{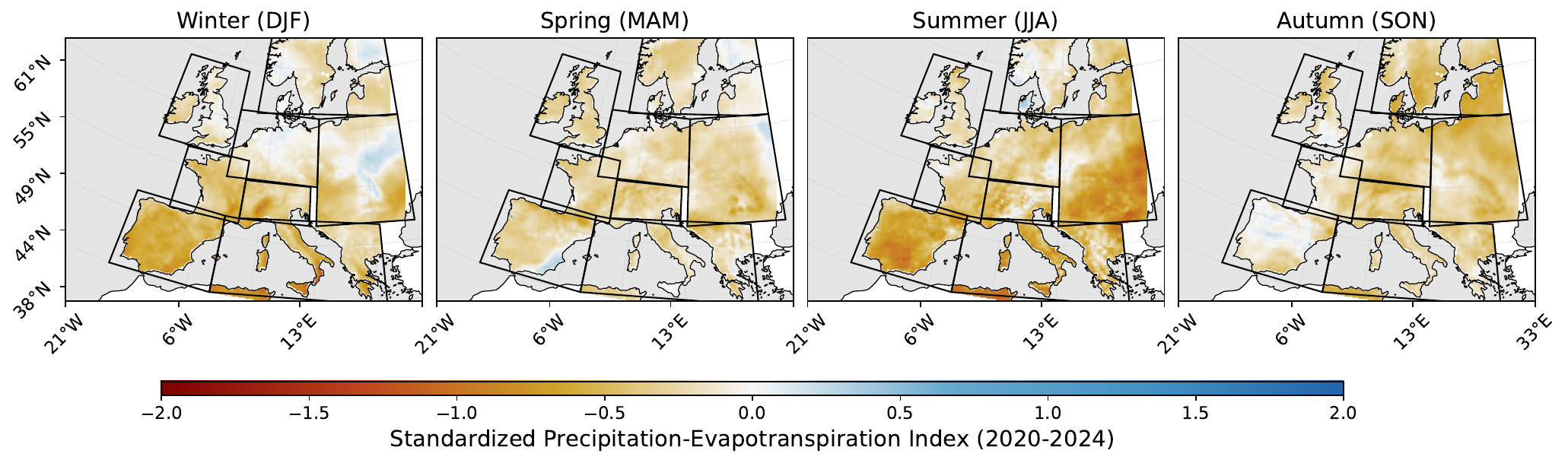}
    \caption{Seasonal SPEI-1 patterns over Europe during the evaluation period 2020--2024.}
    \label{fig:SPEI2020-2024}
\end{figure*}

Machine-learning approaches have shown increasing promise for drought forecasting because they can capture nonlinear relationships between droughts and the hydrometeorological conditions that induce them, thereby partly explaining their variability.
Studies have applied support vector machines \citep{Belayneh2013, Ganguli2014}, neural networks \citep{Morid2007, Dikshit2021}, and hybrid deep-learning architectures including convolutional LSTM variants \citep{DanandehMehr2022, Marusov2024, Ha2024, Vo2023}, with large-scale sea-surface temperature anomalies shown to further improve forecast skill \citep{Li2021}.
However, predictive uncertainty is absent or is inferred solely from the historical record used for model fitting, without explicit representation of the natural variability of the climate \citep{Ganguli2014, Belayneh2016}.

Foundational work established the importance of internal variability for regional uncertainty on shorter timescales \citep{hawkins2009, hawkins2011}.
Recent work has further distinguished aleatoric from epistemic uncertainty and quantified the spatio-temporal contribution of internal variability \citep{GruberFunk2026, gruber2025, deser2020}.
Ensemble prediction systems such as the North American Multimodel Ensemble \citep{Kirtman2014} and ENSEMBLES \citep{Weisheimer2009} estimate forecast uncertainty through the spread across physically consistent members \citep{lavaysse2015, becker2016, Xu2018}.
Yet, the explicit integration of this perspective into AI-based drought forecasting remains limited.

The core contribution of this paper is a method for producing uncertainty-aware drought bounds that explicitly incorporate unexplained internal climate variability.
Furthermore, we show that internal variability of the large ensemble recovers lower-tail drought risk that cannot be learned from a single historical reanalysis record alone.
This perspective is particularly relevant for Europe, where drought risk varies strongly across hydrometeorological regimes and where an ongoing drying trend reflects a shifting climate state (see \Cref{fig:SPEI2020-2024}).
Extreme conditions under new drought regimes stress the reliability of uncertainty estimates and their ability to detect drought, especially when calibrated only on historical data.
Against this background, we present three contributions.
First, we develop a one-month-ahead European drought forecasting framework based on Temporal Fusion Transformer \citep[TFT;][]{Lim2021} forecasts of water balance and a subsequent SPEI-1 transformation.
Second, we extend it to include an uncertainty component that uses large-ensemble climate simulations to quantify spatio-temporal internal variability that cannot be explained by the deep learning algorithm \citep{GruberFunk2026}.
Third, we compare the resulting ensemble-informed lower drought bounds against the prediction quantiles from the historical trajectory.
We evaluate both across European regions and seasons, with attention to practical interpretation as risk-aware drought indicators.
In doing so, the study provides a retrospective proof of concept for uncertainty-aware drought assessment and a transferable framework for climate-variability-aware forecasting.

\section{Methods}

\begin{figure}[h!]
\centering
    \includegraphics[width=\linewidth]{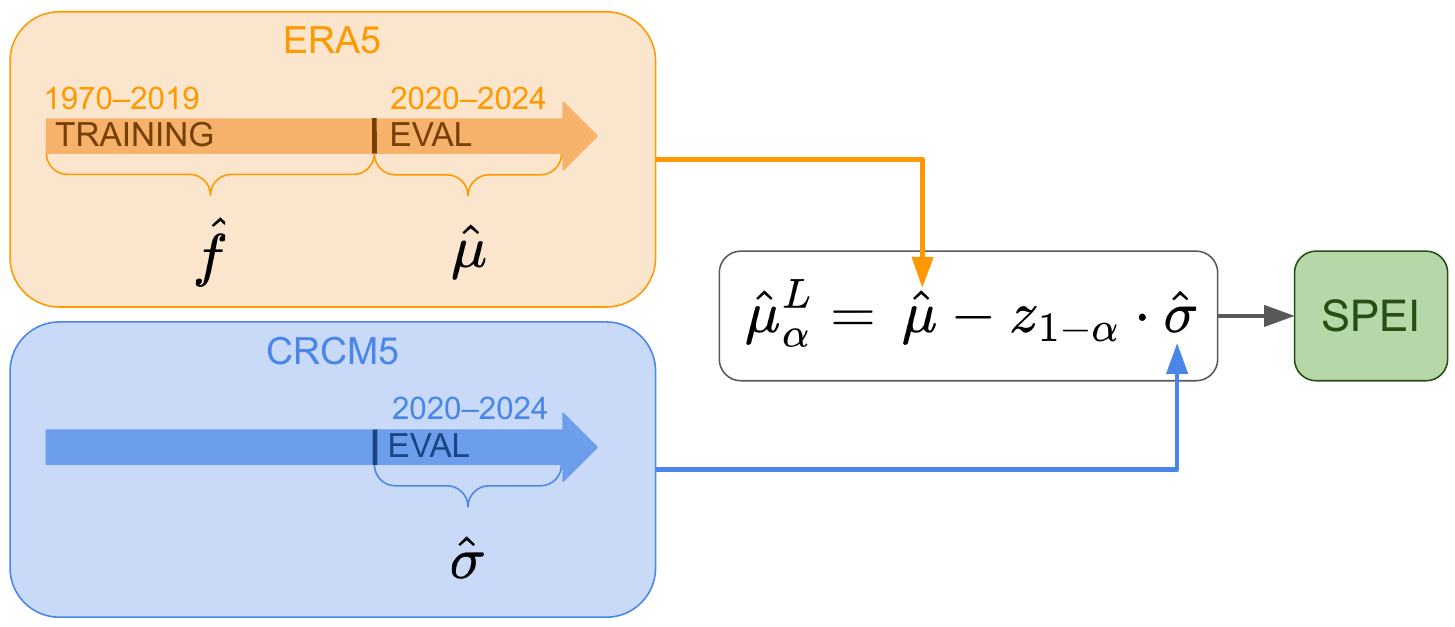}
\caption{Schematic overview of the proposed framework. ERA5-Land data from 1970--2019 are used to train the forecasting model \(\hat{f}\). Applying the trained model to ERA5-Land predictors for 2020--2024 yields the central forecast \(\hat{\mu}\), while applying \(\hat{f}\) to CRCM5-LE simulations over the same period provides an estimate of forecast variability \(\hat{\sigma}\). The resulting forecasts are then transformed to SPEI-1 for evaluation.}
    \label{fig:schema}
\end{figure}

Our aim is to generate monthly European drought forecasts together with uncertainty bounds that reflect internal climate variability and forecast uncertainty.
To this end, we combine a data-driven forecasting model for the expected hydrometeorological trajectory with an ensemble-based estimate of variability obtained from regional climate model simulations.
The workflow is illustrated in \Cref{fig:schema}.
Additionally, we compare against the prediction interval that the forecasting model produces directly from the historical reanalysis.

\subsection{Study Design and Data Basis}

The analysis is conducted for the eight European PRUDENCE regions \citep{christensen2007,bohnisch2021}.
All variables are aggregated to a common monthly temporal resolution and harmonised using bilinear interpolation \citep{schulzweida_2023} to the \(0.11^\circ\) spatial resolution of the climate model before modelling.
Model calibration and training are based on data from 1970 to 2019, while evaluation is carried out on an unseen test period from 2020 to 2024.

We use monthly ERA5-Land reanalysis data as the primary reanalysis dataset for model training \citep{munoz2021,hersbach2020ERA5GlobalReanalysis}.
ERA5-Land provides globally consistent estimates of land-surface and near-surface atmospheric variables from 1950 onward.

To quantify climate variability, we use the Canadian Regional Climate Model Large Ensemble \citep[CRCM5-LE;][]{leduc2019, martynov2013reanalysis, vseparovic2013present}, a single-model initial-condition large ensemble designed for regional analyses of extreme events and internal climate variability \citep{maher2021, funk2025}.
CRCM5-LE dynamically downscales its driving global model, CanESM2, over the European domain.
The ensemble comprises 50 members organised into five families with ten perturbations each.
All members share identical model physics and external forcing, thereby forming a large set of equally plausible, independent climate trajectories \citep{leduc2019, kay2015, thompson2015}.
CRCM5-LE provides a controlled framework for isolating internal climate variability, and its 50-member ensemble enables a statistically robust estimation of its spatio-temporal structure \citep{GruberFunk2026, deser2020, vontrentini2019}.

\subsection{Target Variable and Drought Index}

Our primary hydrometeorological target variable is water balance, defined as the difference between precipitation and potential evapotranspiration \citep{thornthwaite1948}, and is here considered as the monthly average of daily values.
Formally, we denote water balance at location s and time t by \(\mu(s,t)\).

We transform water balance to the SPEI-1 (\citealp{vicente-serrano2010MultiscalarDroughtIndex}).
Negative SPEI-1 values indicate drier-than-normal conditions, whereas positive values indicate wetter-than-normal conditions.
A value below -1.5 indicates drought, values below -2 indicate extreme drought conditions \citep{europeanandglobaldroughtobservatories2025GDOStandardizedPrecipitationEvapotranspiration}.
The SPEI-1 allows for a direct sub-seasonal comparison of drought conditions between regions and time periods \citep{vicente-serrano2010MultiscalarDroughtIndex}.
Details on the calculation of the index are described in \Cref{sec:spei_transform}.

\subsection{Drought Forecasting and Uncertainty Quantification} \label{ssec:method_uq}

\paragraph{Forecasting}
We model the expected evolution of monthly water balance using a Temporal Fusion Transformer \citep{Lim2021}.
A TFT is a deep-learning architecture for time series forecasting with attention-based selection of relevant temporal patterns that outputs both point predictions and quantile estimates.

The TFT is trained and optimised on ERA5-Land data from 1970 to 2019 to learn the relationship between historical hydrometeorological conditions, large-scale predictors, and one-month-ahead local water-balance dynamics.
The model forecasts one-month-ahead water balances for 2020--2024 for each spatial grid cell using lagged target and meteorological variables, large-scale regional predictors (e.g. North Atlantic Oscillation index by \citealp{dawson2016}), calendar information, and static spatial covariates, listed in \Cref{tab:inputs}.
Applying the trained model to ERA5-based predictors yields the central forecast
\begin{equation}
\hat{\mu}(s,t) = \hat{f}\!\left(X(s,t)\right),
\end{equation}
where \(\hat{f}\) denotes the trained forecasting model and \(X(s,t)\) the corresponding ERA5-based predictor set.
This forecast represents the expected water-balance trajectory conditional on the observed large-scale and local hydrometeorological conditions.

\paragraph{Baseline Uncertainty Bound}
Further, we consider a lower water balance bound via the direct quantile estimate,
\begin{equation}
\label{eq:tft_bound}
\hat{\mu}^{RE}_{\alpha}(s,t)= \hat{f}_{\alpha}\!\left(X(s,t)\right),
\end{equation}
where $\alpha$ denotes the quantile level, estimated by the TFT pinball loss, and $RE$ indicates that the lower bound is obtained from reanalysis data.

\begin{table}[t]
\centering
\caption{Input variables used for the drought prediction model.}
\label{tab:inputs}
\footnotesize
\setlength{\tabcolsep}{4pt}
\begin{tabular}{lp{40mm}}
\toprule
Input type & Variable description \\
\midrule
Historical target values
& Water balance \\
\midrule
Historical inputs
& Precipitation \\
 & Temperature above surface \\
& Standardized pressure above surface \\
\midrule
Historical inputs per region
& Sea-surface temperature mean over Mediterranean Sea \\
 & North Atlantic Oscillation index, via EOF \\
\midrule
Time-varying known inputs
& Month \\
& Year \\
\midrule
Static covariates
& Spatial coordinates \\
& Elevation above sea level \\
\bottomrule
\end{tabular}
\end{table}

\paragraph{Ensemble-based Uncertainty Bound}
To quantify uncertainty from internal climate variability, we apply the trained forecasting model \(\hat{f}\) to each CRCM5 ensemble member separately.
This yields a set of trajectories
\begin{equation}
\hat{\mu}_i(s,t) = \hat{f}\!\left(X_i(s,t)\right),
\end{equation}
where \(X_i(s,t)\) denotes the predictor set from ensemble member \(i\).
Now, we consider the residual error of the forecast,
\begin{equation}
    \varepsilon_i(s,t) = {\mu}_i(s,t) - \hat{\mu}_i(s,t),
    \label{eq:eps_i}
\end{equation}
which is the difference between the true water balance $\mu_i(s,t)$ in CRCM5-LE and the forecast $\hat{\mu}_i(s,t)$.
The residual contains the internal variability that remains unexplained after forecasting and the bias induced by applying $\hat{f}$ trained on reanalysis to CRCM5-LE data.
Our aim is to quantify stochastic uncertainty from the unexplained internal climate variability, and the systematic bias is treated as a nuisance component.
Following \cite{GruberFunk2026}, we isolate the unexplained internal variability from the bias through pairwise differences of residuals,
\begin{equation}
\delta_{i,j}(s,t) = \varepsilon_i(s,t)- \varepsilon_j(s,t),
\label{eq:delta_ij}
\end{equation}
for ensemble members \(i\) and \(j\).
Assuming identical bias across members, these differences isolate the component attributable to unexplained internal variability.
We consider squared differences such that the expected value of \Cref{eq:delta_ij} results in
\begin{equation}
E \left[\delta_{i,j}^2(s,t)\right] = 2\sigma^2(s,t),
\end{equation}
which allows us to quantify the internal variability denoted by $\sigma$.
We model the squared pairwise differences as a smooth function of space and time \citep{wood2017}, yielding a location- and time-specific estimate of forecast variability, denoted by \(\hat{\sigma}(s,t)\).
This summarises the expected spread of plausible water-balance trajectories induced by variability, see  \Cref{app:methods} for details.

We combine the central forecast from ERA5-based predictors with the ensemble-based variability estimate to obtain \textit{variability-aware} forecasts.
In particular, for a lower one-sided prediction bound at level \(\alpha\), we define
\begin{equation} \label{eq:CRCM5_lower_bound}
\hat{\mu}^{LE}_{\alpha}(s,t)
=
\hat{\mu}(s,t) - z_{1-\alpha} \cdot \hat{\sigma}(s,t),
\end{equation}
where ${LE}$ indicates the derivation via large ensembles and \(z_{1-\alpha}\) denotes the \((1-\alpha)\)-quantile of the standard normal distribution.
Given the large sample size and monthly temporal aggregation, the central limit theorem ensures that $\hat{\mu}(s,t)$ follows approximately a normal distribution.

Finally, both the central forecast \(\hat{\mu}(s,t)\) and the lower bound \(\hat{\mu}^{LE}_{\alpha}(s,t)\) are transformed from water balance to SPEI-1 units.
The method is described in more detail in \Cref{sec:spei_transform}.
This yields a central drought forecast together with a lower SPEI-1 bound that can be interpreted as a risk-aware estimate of future drought severity.

\paragraph{Interpretation} A lower bound represents a physically plausible worst-case trajectory of future drought indication, quantifying how far conditions could deteriorate below the expected forecast.
The lower bound in \Cref{eq:tft_bound} is obtained directly from the predictive quantile trained on the single historical ERA5-Land trajectory and represents the uncertainty that can be learned from reanalysis data.
In this paper, we derived a lower bound based on the CRCM5 large ensemble in \Cref{eq:CRCM5_lower_bound}.
It exploits climate model information from the 50 large-ensemble trajectories to estimate a more stable bound on internal variability.

\section{Results}

The results are organised into two levels of spatial aggregation.
\Cref{sec:regional} evaluates drought detection at the regional level.
The SPEI-1 is calculated from the monthly mean water balance in each of the eight PRUDENCE regions, and the resulting index reflects large-scale coherent hydrometeorological anomalies.
\Cref{sec:seasonal} evaluates a local SPEI-1 at grid cells of 0.11$^\circ$ spatial resolution for each calendar season.
This preserves the spatial detail needed to assess the bounds across the heterogeneous European climate and within each season.
In both analyses, the evaluation is restricted to the out-of-sample test period 2020--2024.
The model uses the variables listed in \Cref{tab:inputs} as input data in the form of a 12 or 24 month time series up to the month preceding the forecast.
We compare two lower bounds introduced in \Cref{ssec:method_uq}: one derived from a single ERA5-Land reanalysis trajectory (hereafter \textit{Reanalysis}), and one from the 50-member CRCM5 large ensemble (hereafter \textit{Large Ensemble}).
Bound quality is assessed by the fraction of observed ERA5-Land SPEI-1 values that fall \emph{below} each nominal 10\% lower bound and, critically, at the drought and extreme-drought thresholds, where reliable lower-tail coverage matters most for risk assessment.

\subsection{Regional SPEI-1}\label{sec:regional}
\subsubsection*{Forecast performance}

The forecasting error, measured by the mean absolute error (MAE, \citealp{willmott2005AdvantagesMeanAbsolute}), is reported in \Cref{tab:region_metrics_test} for the test period.
\Cref{tab:region_metrics_full} in the Appendix provides region-level performance metrics for the full calibration and evaluation period 1970--2024, serving as a reference baseline against the test period results.
The MAE spans from 0.42 to 1.11, representing varying levels of forecast skill.
The Iberian Peninsula and the Mediterranean are the best-performing domains reflecting the strong precipitation seasonality of the Mediterranean climate.
The Alps exhibit the weakest forecast skill given complex precipitation patterns and high precipitation variability in this mountainous region.
Also, the British Isles show limited skill, reflecting the strong dependence of regional precipitation on large-scale circulation variability beyond what the NAO index alone captures.

\begin{figure*}[p]
    \centering
    \includegraphics[width=.9\linewidth]{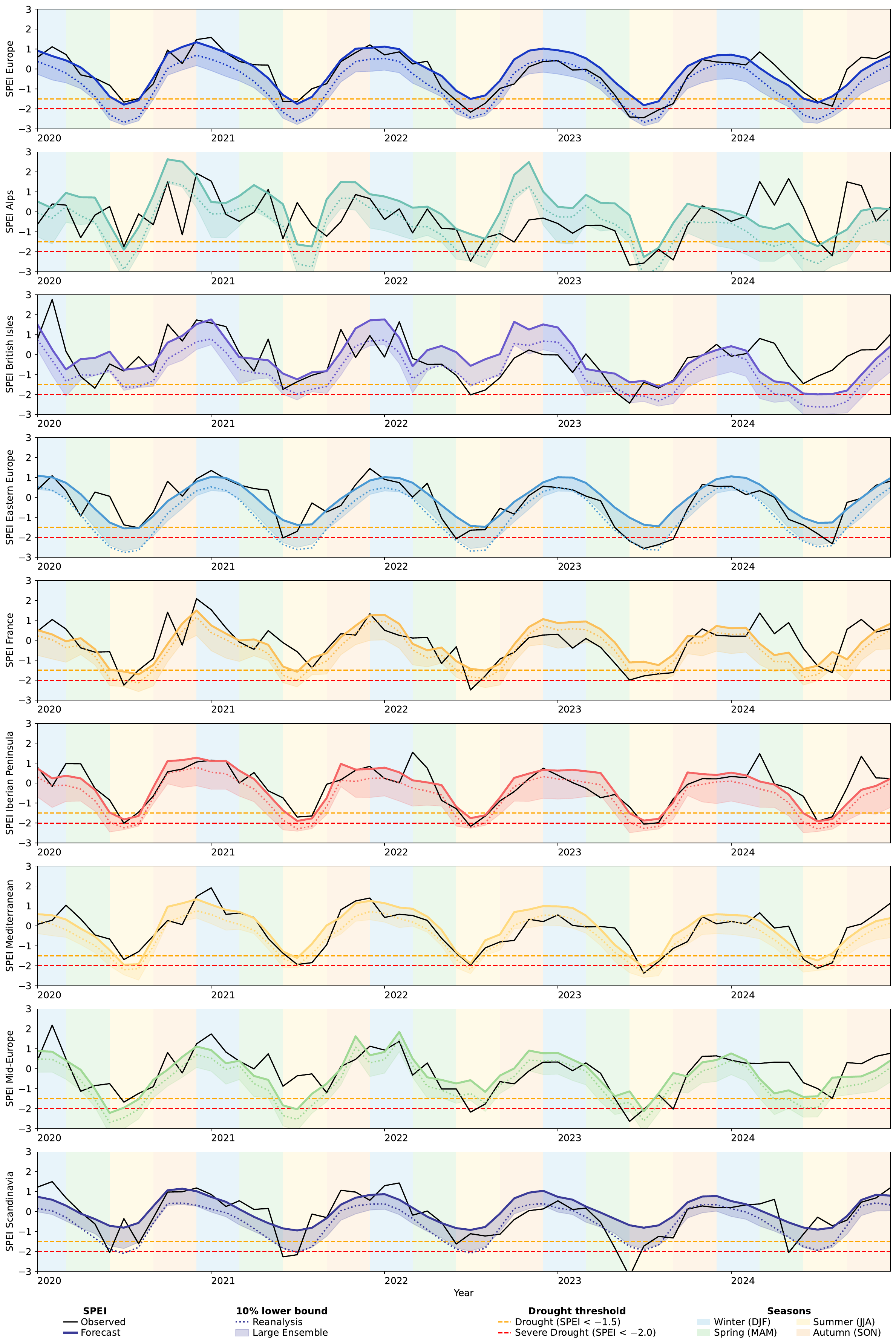}
    \caption{Prediction overview of the SPEI-1 for the entire domain of Europe and the eight European regions during the evaluation period 2020--2024 (solid colored lines).
    Observed ERA5-Land SPEI-1 (black line) is shown against forecasted values and 10\% uncertainty bounds.
    Dotted bounds are reanalysis-based TFT quantiles; shaded bounds are CRCM5 ensemble-based bounds.
    Background colouring indicates the season.}
    \label{fig:regional_spei_eu}
\end{figure*}

Comparing the observed and forecast trajectories in \Cref{fig:regional_spei_eu}, the SPEI-1 forecast for Europe and its regions indicates that the observed drying trend of 2022--2023 is underestimated. This is consistent with the tendency of reanalysis-trained models to regress toward average historical climatology and demonstrates the need for informative lower bounds on drought forecasts.

\begin{table}[t]
\centering
\caption{Region-level performance metrics for the test period 2020--2024.
MAE is computed on the median SPEI-1 forecast against ERA5-Land.
The two rightmost columns show the percentage of observed SPEI-1 values that fall \emph{below} the respective 10\% lower bound, i.e.\ the share of observations not captured by the bound.
The reanalysis bound is the TFT direct quantile learned from ERA5-Land alone; the large-ensemble bound additionally uses CRCM5 ensemble variability.
A well-calibrated 10\% lower bound yields values close to 10\%.}
\label{tab:region_metrics_test}
\footnotesize
\setlength{\tabcolsep}{4pt}
\begin{tabular}{lrrr}
\toprule
& \multicolumn{1}{c}{\textbf{Forecast error}} & \multicolumn{2}{c}{\textbf{\% below 10\% bound}} \\
\cmidrule(lr){2-2}\cmidrule(lr){3-4}
\textbf{Region} & \textbf{MAE} & \textbf{Large Ensemble} & \textbf{Reanalysis} \\
\midrule
Alps                 & 1.11 & 23.33 & 46.67 \\
British Isles        & 0.82 & 23.33 & 28.33 \\
Eastern Europe       & 0.50 & 13.33 & 10.00 \\
France               & 0.66 & 13.33 & 43.33 \\
Iberian Peninsula    & 0.42 &  1.67 & 15.00 \\
Mediterranean        & 0.44 &  6.67 & 26.67 \\
Mid-Europe           & 0.64 & 11.67 & 26.67 \\
Scandinavia          & 0.57 & 20.00 & 23.33 \\
\bottomrule
\end{tabular}
\end{table}

\subsubsection*{Uncertainty bounds}
The reanalysis-based lower bound is inadequately calibrated, exhibiting bound exceedance far over 10\%  in \Cref{tab:region_metrics_test}.
In the Alps, 46.67\% of observations fall below this bound, and values above 25\% are seen in the British Isles, France, the Mediterranean and Mid-Europe.
The poor calibration can also be seen in \Cref{fig:regional_spei_eu}, which shows that reanalysis bounds fail to cover many observed SPEI-1 values.
In France, the reanalysis bound captures only 1 of 8 drought events and 1 of 2 extreme drought events; in the Mediterranean, it captures only 5 of 9 drought events.
Only Eastern Europe, where the bound is near the nominal level of 10\%, constitutes an exception.

The ensemble-based lower bound is better calibrated in most regions.
Six of the eight domains show exceedance at or below 20\% (see \Cref{tab:region_metrics_test}), and at the pan-European level, it captures all drought events during the test period (see \Cref{fig:regional_spei_eu}).
The Iberian Peninsula and the Mediterranean, which benefit from strong precipitation seasonality, show the best calibration for extremes, as the ensemble bounds cover all drought and extreme drought events in both regions.
In the Alps, the ensemble bound covers only every second drought event, consistent with the weakest forecast skill in the domain and the unresolved orographic variability noted above.
The British Isles' strong dependence on circulation-driven moisture deficits \citep{West2021} is not fully represented in the ensemble spread, constraining detectability.

The ensemble bound closes much of this gap, but it remains anchored to the central TFT forecast. Where the model itself struggles, as in the Alps and the British Isles, the bound's ability to detect extremes is also weakened.
The advantage of the large-ensemble bound over the reanalysis bound lies in its robustness, due to a larger sample size, and in its ability to adapt the uncertainty width to shifting future climate states rather than being fixed to historical variability alone.

\subsection{Seasonal SPEI-1}\label{sec:seasonal}

\subsubsection*{Forecast performance}

\begin{table}[t]
\centering
\caption{Seasonal SPEI-1 performance metrics for the test period 2020--2024, grouped by season based on reanalysis.
MAE is computed on the expected SPEI-1 forecast against ERA5-Land reanalysis.
The two rightmost columns show the percentage of observed SPEI-1 values that fall \emph{below} the respective 10\% lower bound.
The reanalysis bound is the direct TFT quantile; the large-ensemble bound incorporates CRCM5 ensemble variability.
A well-calibrated 10\% lower bound should yield values close to 10\% over all events.}
\label{tab:pixel_metrics_seasonal_alps}
\footnotesize
\setlength{\tabcolsep}{4pt}
\begin{tabular}{lrrr}
\toprule
& \multicolumn{1}{c}{\textbf{Forecasting error}} & \multicolumn{2}{c}{\textbf{\% below 10\% bound}} \\
\cmidrule(lr){2-2}\cmidrule(lr){3-4}
\textbf{Season} & \textbf{MAE} & \textbf{Large Ensemble} & \textbf{Reanalysis} \\
\midrule
Winter (DJF) & 0.93 & 11.8 & 33.4 \\
Spring (MAM) & 0.90 & 15.5 & 22.1 \\
Summer (JJA) & 1.01 & 24.3 & 28.7 \\
Autumn (SON) & 0.80 & 20.0 & 36.9 \\
\textbf{All Seasons} & \textbf{0.908} & \textbf{17.9} & \textbf{30.3} \\
\bottomrule
\end{tabular}
\end{table}

At 0.11$^\circ$ resolution, the pan-European MAE is 0.908 across all grid cells and seasons, reflecting the greater difficulty of local variability patterns (\Cref{tab:pixel_metrics_seasonal_alps}).
The seasonal forecast error varies.
The best performance is achieved in autumn, while the forecast in summer is the most demanding.
The skill deficit in summer is consistent with the dominance of convective precipitation over central Europe, which introduces high local variability that is difficult to capture at monthly lead times.

\subsubsection*{Uncertainty bounds across space and season}

\begin{figure*}[t]
    \centering
    \includegraphics[width=\linewidth, trim=0 65 0 0, clip]{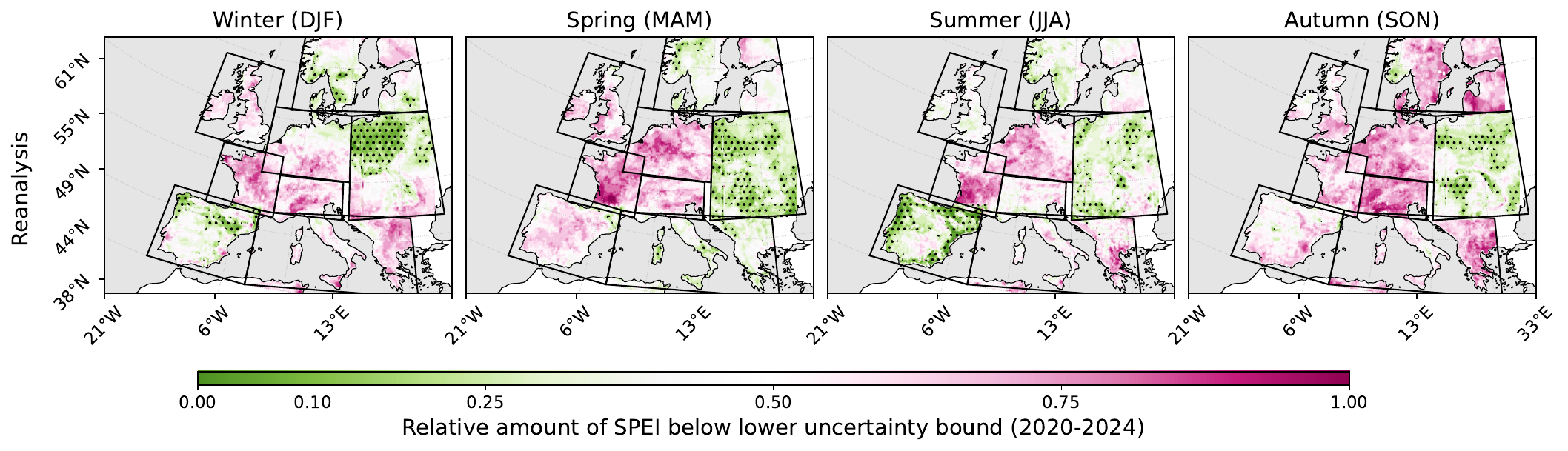}
    \vspace{-1mm}
    \includegraphics[width=\linewidth, trim=0 0 0 22, clip]{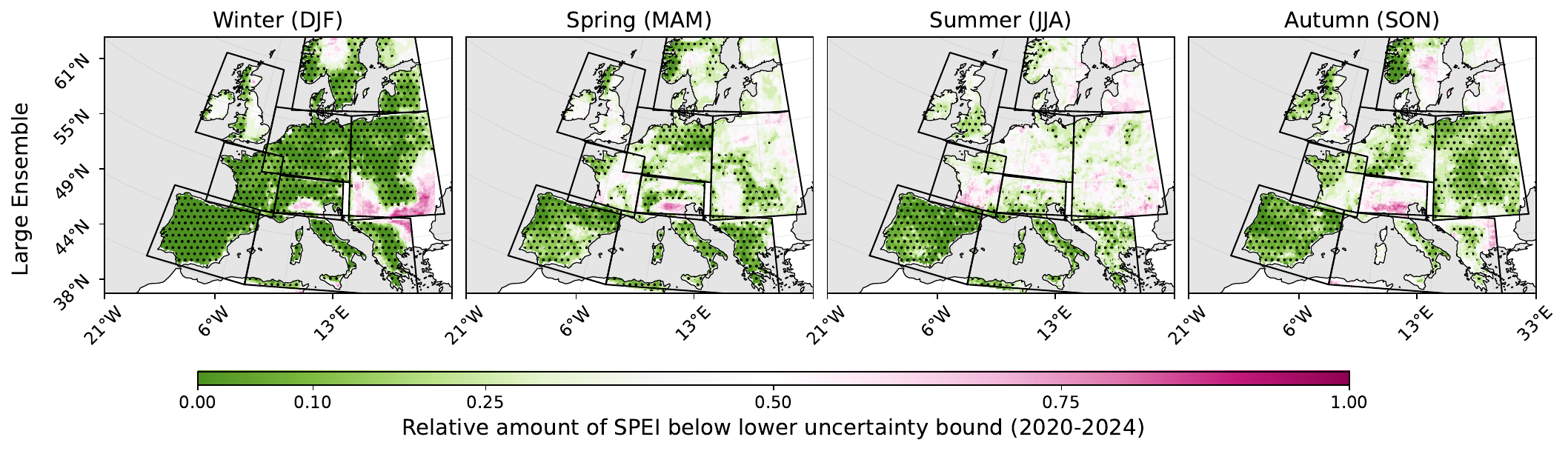}
    \caption{Fraction of observed seasonal SPEI-1 values falling below the 10\% lower reanalysis bound (top row) and ensemble bound (bottom row), by season and grid cell, for the period 2020--2024.
    Values close to 10\% (green) indicate a well-calibrated bound; values well above 10\% indicate that the bound underestimates drought risk and many drought observations are not captured (pink).
    Hatching marks grid cells with exceedance between 0 and 20\%.}
    \label{fig:SPEI2020-2024_relative}
\end{figure*}

In \Cref{fig:SPEI2020-2024_relative}, we investigate the capability of both bounds to capture drought events for each location in Europe for each season.
The lower bound based on reanalysis is poorly calibrated in most of continental Europe.
Near-ideal detection rates, indicated by the hatched area in \Cref{fig:SPEI2020-2024_relative}, are achieved only partially over Eastern Europe and in limited parts of Scandinavia, and only in winter and summer over the Iberian Peninsula.
Pan-European exceedance averages 30.3\% across all grid cells and seasons, that is three times the ideal level of 10\% (\Cref{tab:pixel_metrics_seasonal_alps}, last column).

The ensemble-based bound (\Cref{fig:SPEI2020-2024_relative}, bottom row) covers a substantially larger area in every season.
Southern Europe is near-perfectly calibrated in all seasons, with detection rates close to the ideal 10\% almost everywhere.
Most of Europe achieves near-ideal calibration, with the main exception being the southeastern mountain ranges, including the Dinaric Alps and the Carpathians, as well as the Alps, where complex local orographic variability is beyond the capacity of CRCM5-LE at a 12km resolution.
Spring and autumn show broadly similar spatial patterns, with good calibration across southern and central Europe but persistent underperformance over the Alps.
Winter is the overall best-performing season with only 11.8\% pan-European exceedance of SPEI-1 values (\Cref{tab:pixel_metrics_seasonal_alps}).
Summer is the most challenging season with an exceedance of 24.3\% against the nominal 10\%.
Taken together, the ensemble bound improves substantially on the reanalysis bound, which fails in large parts of the domain.

The contrast between the two bounds increases further under drought conditions.
The reanalysis-based bound fails to detect 74.2\% of drought events.
The ensemble-based bound fails to detect every second drought, with winter performing substantially better than summer.
Under extreme drought, the reanalysis bound performs worst: 90.3\% of extreme drought observations fall below the bound overall.
The ensemble-based bound performs better overall, with 71.4\% of extreme drought events not captured.
This progressive degradation with increasing drought severity reflects the inherent difficulty of modelling rare lower-tail events reliably. The reanalysis bound shows additional structural limitations that regress toward historically typical tail widths and cannot account for the anomalous conditions of 2020--2024.
The ensemble lower bound retains a realistic width in the lower tail precisely where calibration is most consequential for risk assessment.

\section{Discussion}
Subseasonal drought forecasting sits in what \citet{richter2025} and \citet{guan2026} call the predictability desert, where internal variability dominates and uncertainty is routinely underestimated.
Our results suggest that the large-ensemble approach does not escape that desert but partially maps it: the ensemble-based bound quantifies the internally generated spread that a single reanalysis trajectory cannot recover, improving lower-tail coverage across most of Europe during a historically anomalous test period.
This agrees with earlier work showing that machine-learning models can capture nonlinear drought-relevant relationships \citep{Belayneh2013, Ganguli2014, Dikshit2021, DanandehMehr2022, Li2021, Marusov2024}, while demonstrating that point forecasting alone is insufficient.

The years 2020--2024 include drought conditions that exceed the ERA5-Land record.
In this period, the reanalysis lower bound regressed toward historically typical tail widths.
The ensemble-based bound draws on 50 physically consistent trajectories and isolates internal variability through pairwise residual differences.
This connects the large-ensemble logic of probabilistic early warning \citep{Weisheimer2009, Kirtman2014, lavaysse2015, becker2016, Xu2018} to a reanalysis-trained machine-learning predictor.

Compared to the reanalysis, the ensemble bound remains more conservative when extremes are physically plausible.
This conservatism comes at the cost of sharpness, reflecting a trade-off between extreme-drought detectability and predictive sharpness.
This trade-off can be controlled directly by the bound level $\alpha$: the nominal 10\% bound stays sharp and well calibrated for overall drought assessment, while tightening to 5\% or 2.5\% widens the lower tail to improve detection of drought and extreme drought where the reanalysis bound collapses.

In the Alps, the 12\,km ensemble resolution cannot resolve local orographic precipitation patterns, so the estimated variability underrepresents the true spread of plausible water-balance trajectories in complex terrain.
The resulting high extreme-drought exceedance rates suggest that lower bounds at the 5\% or 2.5\% levels may be more appropriate for detecting more extreme events.
The NAO index captures only part of the large-scale atmospheric variability, leaving a component of forecast uncertainty unrepresented in the bounds.
Finally, a further advantage of ensemble-based bounds is that they enable estimates of how internal variability shifts under global warming as the CRCM5-LE simulations extend into the future.

\section{Conclusion}

This study introduces an uncertainty-aware framework for European drought prediction that combines a Temporal Fusion Transformer with large-ensemble estimates of internal climate variability.
Its novelty lies in using climate-model ensemble simulations not as direct observations, but as a means of learning the scale and spatio-temporal structure of internally generated forecast variability.
This allows historical and future simulations to inform uncertainty estimation while ERA5-Land remains the verification reference.
The resulting lower SPEI-1 bounds recover drought tail risk more reliably than the TFT's direct quantile estimates, especially during the anomalously dry 2020--2024 test period, precisely because they are not constrained to the variability expressed in a single historical trajectory.

The main significance of the work is therefore conceptual as well as practical: internal variability is treated as a forecast quantity in its own right rather than as irreducible noise.
The framework is transferable to other regions, drought indices, accumulation windows, and multistep lead times.
Future work should extend the approach to multi-model ensembles to reduce sensitivity to driving-model biases, evaluate non-Gaussian tail models for compound-event seasons, include additional land-surface and snow-process predictors to address the Alpine and British Isles failure modes, and test integration with operational seasonal prediction systems.

\section*{Funding}
CG is supported by the DAAD programme Konrad Zuse Schools of Excellence in Artificial Intelligence, sponsored by the Federal Ministry of Research, Technology and Space.

\section*{Conflict of Interest}

The authors declare that they have no competing financial interests or personal relationships that could have appeared to influence the work reported in this paper.

\section*{Author contributions}

\noindent Henri Funk \orcid{0009-0007-0949-8385} -- Conceptualization (equal), Data curation (lead), Formal analysis (lead), Investigation (lead), Methodology (equal), Software (equal), Validation (equal), Visualization (lead), Writing – original draft (equal), Writing – review \& editing (equal)

\medskip

\noindent Cornelia Gruber \orcid{0009-0002-0657-3558} -- Conceptualization (equal), Formal analysis (supporting), Investigation (supporting), Methodology (equal), Software (equal), Validation (equal), Visualization (supporting), Writing – original draft (equal), Writing – review \& editing (equal)

\medskip

\noindent Göran Kauermann \orcid{0000-0003-0742-7835} -- Supervision (equal), Conceptualization (equal), Methodology (equal), Writing – review \& editing (equal)

\medskip

\noindent Helmut Küchenhoff \orcid{0000-0002-6372-2487} -- Supervision (equal), Conceptualization (equal), Methodology (equal), Writing – review \& editing (equal)

\medskip

\noindent Magdalena Mittermeier \orcid{0000-0002-8668-281X} -- Supervision (equal), Conceptualization (equal), Methodology (equal), Writing – review \& editing (equal)

\section*{Data availability}
The ERA5-Land reanalysis data that support the findings of this study are openly available in the Climate Data Store of Copernicus at \url{https://doi.org/10.24381/cds.68d2bb30}. The CRCM5-LE data for the historical and RCP8.5 simulations are available from the ClimEx project at \url{https://www.climex-project.org/data-access/}. The processed data that support the findings of this study are available upon reasonable request from the authors.

\bibliographystyle{abbrvnat}
\bibliography{references}

\appendix
\section{Methodological details}

\subsection{Derivation of Method}
\label{app:methods}
We now go into more detail on how the method extracts spatio-temporal internal variability. For further details, we refer to \cite{GruberFunk2026}.
The residual $\varepsilon_i$ defined in \Cref{eq:eps_i} consists of systematic parts that come from the bias that is induced by applying $\hat{f}$, a model trained on ERA5, to a new domain, CRCM5 climate simulations, denoted by $\psi$. It further consists of all remaining variability that is not captured by the forecasting model, like the climate's internal variability, denoted by $\vartheta_i$.
\begin{equation}
    \varepsilon_i(s,t) = \mu_i(s,t) - \hat{\mu}_i(s,t) = \psi(s,t) + \vartheta_i(s,t)
\end{equation}
Since the climate members follow the identical data simulation protocol and all variation comes from small perturbations in atmospheric conditions, and the identical forecasting model $\hat{f}$ is used, it is reasonable to assume that the systematic bias $\psi$ is identical across residuals of various climate members, $\varepsilon_i$ and $\varepsilon_j$. It thus cancels when calculating differences.
\begin{align}
   \delta_{i,j}(s,t) &= \varepsilon_i(s,t) - \varepsilon_j(s,t) \nonumber\\
   &= \psi(s,t) + \vartheta_i(s,t) - \left[\psi(s,t) + \vartheta_j(s,t) \right] \nonumber\\
   &= \vartheta_i(s,t) - \vartheta_j(s,t)
\end{align}
Now with the light assumptions, that each $\vartheta_i$ has a mean of zero, and is independent with equal variance across all members $i, j$:
\begin{align} \label{eq:nat_var_true}
    E[\delta_{i,j}^2(s,t)] &= E[(\vartheta_{i}(s,t) - \vartheta_{j}(s,t))^2] \nonumber\\
    &= E[\vartheta_{i}^2(s,t)] - 2E[\vartheta_{i}(s,t)\vartheta_{j}(s,t)] + E[\vartheta_{j}^2(s,t)] \nonumber\\
    &= \sigma^2(s,t) + \sigma^2(s,t) = 2\sigma^2(s,t).
\end{align}
This means, that modelling the expected value of $\delta_{i,j}^2$ gives a direct estimate of two times the internal variability $\sigma^2$.

\subsection{Statistical model}
We model the squared differences using a generalized additive model:
\[
\delta_{i,j}^2(s,t) \sim \text{Gamma}(\mu_{i,j}(s,t), \phi),
\]
with log link
\begin{align*}
\log \mu_{i,j}(s,t) = {}& \beta_0 + \beta_1 \cdot \text{year}
+ f_1(\text{month}) \\
& + f_{\text{season}}(\text{lon}, \text{lat})
+ u_{i,j}.
\end{align*}

Here, $\mu_{i,j}(s,t) = \mathbb{E}[\delta_{i,j}^2(s,t)]$, and the smooth functions $f_k$ are represented using penalized splines and $u_{i,j}$ a random intercept for the member pair.
Spatial effects are modeled using tensor product smooths, and a cyclic spline is used for the seasonal component.
Transforming the estimates gives us the standard deviation, capturing the internal variability:
\begin{equation} \label{eq:sigma_theoretical}
    \hat{\sigma}(s,t) = \sqrt{\frac{1}{2} \; {\hat{\mu}_{i,j}(s,t)}}.
\end{equation}

This yields space and time specific estimates of uncertainty, $\hat{\sigma}(s,t)$, which we use in \Cref{eq:CRCM5_lower_bound}.

\subsection{SPEI-1 transformation}
\label{sec:spei_transform}

Both the expected forecast $\hat{\mu}(s,t)$ and the lower bound $\hat{\mu}^{LE}_\alpha(s,t)$ are expressed in units of water balance.
For drought assessment, the water balance is transformed to the SPEI-1 following \citet{vicente-serrano2010MultiscalarDroughtIndex}.
The SPEI-1 is calibrated by fitting a three-parameter Gamma to the ERA5-Land water-balance series over the calibration period 1970--2019 via maximum likelihood,
\begin{equation}
    \hat{\theta}(s) = \operatorname*{arg\,max}_{\theta}\sum_{t \in \mathcal{T}_{\mathrm{cal}}}
    \log f_\Gamma\!\left(\mu(s,t);\,\theta\right),
    \label{eq:spei_fit}
\end{equation}
where $f_\Gamma$ is the Gamma density and $\mathcal{T}_{\mathrm{cal}}$ denotes the set of calibration time steps.
Water balance is transformed to the SPEI-1 by
\begin{equation}
    \mathrm{SPEI\text{-}1}(x, s) = \Phi^{-1}\!\left(F_\Gamma\left(x;\,\hat{\theta}(s)\right)\right),
    \label{eq:spei_transform}
\end{equation}
where $F_\Gamma(\cdot;\hat{\theta}(s))$ is the cumulative distribution function fitted at location $s$, $\Phi^{-1}$ the standard-normal quantile function.
Applying this transformation to all out-of-sample quantities (2020--2024) yields directly interpretable drought severity estimates for the observed, forecast, and uncertainty-bound trajectories on a common scale.

\subsection{Hyperparameters}
\label{sec:app_methods}

\begin{table*}[!t]
\centering
\caption{Setup and optimal hyperparameter configurations of the Temporal Fusion Transformer for each European subdomain. \textit{Locations} denotes the number of grid cells within each domain.
The remaining entries correspond to hyperparameters that govern batch loading, model architecture, and optimisation.
\textit{Normalisation} indicates whether the target variable was standardised per grid cell (\textit{loc}) or across the entire subdomain (\textit{all}).
\textit{Optimizer} specifies the optimization algorithm used, either Adam (\textit{adam}) or decoupled weight-decay Adam (\textit{adamw}).
The  \textbf{test loss} reports the quantile loss (for $\tau = 0.1, 0.5, 0.9$) evaluated on the independent holdout period 2020--2024.
}
\label{tab:tft_hyperparams}
\small
\setlength{\tabcolsep}{5pt}
\begin{tabular}{lcccccccc}
\toprule
 & \textbf{al} & \textbf{bi} & \textbf{ea} & \textbf{fr} & \textbf{ip} & \textbf{md} & \textbf{me} & \textbf{sc} \\
\midrule
\textit{Locations} & 2016 & 1820 & 7887 & 2046 & 4147 & 3556 & 3903 & 6121 \\
\textit{Normalization} & all & all & loc & all & loc & loc & all & all \\
\textit{Batch size} & 512 & 2048 & 1024 & 512 & 256 & 1024 & 32 & 1024 \\
\textit{Encoder length} & 12 & 12 & 24 & 12 & 24 & 12 & 12 & 12 \\
\textit{Hidden continuous size} & 8 & 32 & 16 & 16 & 32 & 64 & 32 & 64 \\
\textit{Hidden size} & 512 & 128 & 32 & 512 & 128 & 128 & 128 & 64 \\
\textit{LSTM layers} & 2 & 2 & 2 & 4 & 2 & 4 & 2 & 2 \\
\textit{Attention head size} & 4 & 4 & 2 & 8 & 4 & 8 & 8 & 8 \\
\textit{Dropout} & 0.07 & 0.18 & 0.00 & 0.19 & 0.15 & 0.17 & 0.23 & 0.16 \\
\textit{Gradient clip value} & 10.00 & 1.00 & 10.00 & 0.00 & 1.00 & 0.10 & 0.00 & 0.10 \\
\textit{Learning Rate} & $1\times10^{-5}$ & $3\times10^{-6}$ & $1\times10^{-5}$ & $5\times10^{-6}$ & $1\times10^{-5}$ & $3\times10^{-6}$ & $1\times10^{-5}$ & $5\times10^{-6}$ \\
\textit{Optimizer} & adamw & adam & adamw & adam & adam & adamw & adamw & adam \\
\textbf{Test Loss} & 1.00 & 0.42 & 0.26 & 0.36 & 0.46 & 0.67 & 0.32 & 0.64 \\
\bottomrule
\end{tabular}
\end{table*}

The hyperparameter space is depicted in \Cref{tab:hp}. \Cref{tab:tft_hyperparams} specifies details on data and training for the 8 European subdomains.

\begin{table}[!t]
\centering
\caption{Hyperparameter search space for Temporal Fusion Transformer (TFT) model.
Continuous hyperparameters are sampled uniformly or log-uniformly, as indicated.
Conditional parameters (e.g., gradient accumulation) are activated only for specific batch sizes.}
\label{tab:hp}
\tiny
\begin{tabular}{lll}
\toprule
\textbf{Category} & \textbf{Hyperparameter} & \textbf{Search space / Values} \\
\midrule
\multirow{4}{*}{Training schedule}
 & Batch size & $\{32,\;256,\;1024\}$ \\
 & Gradient accumulation (large batch) & $\{1,\;2,\;4\}$ (only if batch size $=1024$) \\
 & Training batch limit (default) & $\{10,\;25,\;100\}$ (only if batch size $\neq1024$) \\
 & Gradient clipping & $\{0.001,\;0.1,\;1.0,\;10.0\}$ \\
\midrule
\multirow{7}{*}{Model architecture}
 & Max. encoder length & $\{12,\;24\}$ \\
 & Normalization strategy & \texttt{loc}, \texttt{all} \\
 & Dropout rate & Uniform$(0,\;0.3)$ \\
 & LSTM layers & $\{1,\;2,\;4\}$ \\
 & Attention head size & $\{2,\;4,\;8\}$ \\
 & Hidden size & $\{16,\;32,\;64,\;128\}$ \\
 & Continuous hidden size & $\{8,\;16,\;32,\;64\}$ \\
\midrule
\multirow{2}{*}{Optimization}
 & Optimizer & \texttt{Adam}, \texttt{AdamW} \\
 & Learning rate & Log-uniform$(10^{-5},\;5\times10^{-3})$ \\
\midrule
\multirow{2}{*}{Search configuration}
 & Number of trials & 300 \\
 & Time budget & 29 hours \\
\bottomrule
\end{tabular}
\end{table}

\subsection{Full-period performance metrics}

\begin{table}[h!]
\centering
\caption{Region-level performance metrics for the full period 1970--2024.
Columns as in \Cref{tab:region_metrics_test}.}
\label{tab:region_metrics_full}
\footnotesize
\setlength{\tabcolsep}{4pt}
\begin{tabular}{lrrr}
\toprule
& \multicolumn{1}{c}{\textbf{Forecast error}} & \multicolumn{2}{c}{\textbf{\% below 10\% bound}} \\
\cmidrule(lr){2-2}\cmidrule(lr){3-4}
\textbf{Region} & \textbf{MAE} & \textbf{Large Ensemble} & \textbf{Reanalysis} \\
\midrule
\textbf{Europe}      & \textbf{0.25} & \textbf{0.26} & \textbf{2.60} \\
\midrule
Alps                  & 0.60 & 7.07 & 14.65 \\
British Isles         & 0.45 & 5.81 &  8.84 \\
Eastern Europe        & 0.42 & 3.91 &  3.39 \\
France                & 0.40 & 5.56 & 19.19 \\
Iberian Peninsula     & 0.25 & 0.52 &  5.47 \\
Mediterranean         & 0.38 & 5.73 & 13.28 \\
Mid-Europe            & 0.41 & 5.05 & 12.12 \\
Scandinavia           & 0.44 & 7.32 &  7.07 \\
\bottomrule
\end{tabular}
\end{table}

The TFT is optimized for a 1-month horizon (1-month lead time) and a max encoder length (input) of 1 to 2 years.
Water balance is either standardized over the entire region (\texttt{all}) or location-wise (\texttt{loc}).
To compare the prediction of lower quantiles, we include pinball loss that predicts the quantiles 0.1, 0.5, and 0.9
We train and validate the model on different hyperparameter settings using 50 years from 1970 to 2019.
All our results are evaluated on an unseen test set from 2020 to 2024.

\end{document}